\documentclass[slac_one]{revtex4}
\usepackage{graphicx}
\usepackage{fancyhdr}
\begin{document}

\title{Measurement of the Inclusive Isolated Photon Cross Section at CDF} 

%

\author{C. Deluca, M. Mart{\'\i}nez}
\affiliation{Institut de F{\'\i}sica d'Altes Energies (IFAE), Universitat Aut{\`o}noma de Barcelona,
Spain}
\author{R. Culbertson, S.-S. Yu}
\affiliation{FNAL, Batavia, IL 60510, USA}

\author{For the CDF Collaboration}

\begin{abstract}
We present preliminary results on inclusive direct photon production in $p\bar{p}$ collisions at
$\sqrt{s}$=1.96~TeV, using data collected with the upgraded Collider Detector at Fermilab in Run II,
corresponding to an integrated luminosity of 451 pb$^{-1}$. Measurements are performed as a
function of the photon transverse momentum for photons with $p_{T}>$30~GeV and $|\eta|<$1.0. Photons
are required to be isolated in the calorimeter. The measurement is corrected to
the hadron level and compared to NLO pQCD predictions. 
\end{abstract}

\maketitle

\thispagestyle{fancy}


\section{INTRODUCTION}

Prompt photons can be produced via two possible mechanisms, directly from the hard subprocess,
or as a result of the collinear fragmentation of a parton that is itself produced with a large
transverse momentum. The tree level contributions to the hard subprocess are the annihilation
($q\bar{q}\rightarrow g\gamma$) and the Compton ($q(\bar{q})g\rightarrow q(\bar{q})\gamma$) processes.

The production of inclusive prompt photon provides a stringent test of pQCD
predictions. Its measurement offers several advantages with respect to pure QCD processes such as
inclusive production of jets \cite{bib:CTEQ}. First, the presence of the QED vertex at tree level 
makes the theory calculations more reliable. The process also gives access to lower $p_{T}$'s than
jets, where the underlying event contamination reduces the sensitivity
of the QCD measurements. Photons do not hadronize, so there is no need for arbitrary jet definitions, and
the photon energies can be measured with electromagnetic rather than hadronic calorimeters, resulting in
improved energy resolution. 

One of the main motivations for prompt photon measurements is their potential to constrain
the gluon distribution of the proton. This is due to the gluon appearance
in the initial state of the tree level Compton diagrams, which  dominate the photon cross
section at low-to-moderate $p_T$.



The main background contamination to the direct photon production comes from photons from meson
decays (mostly $\pi^{0}$'s and $\eta$'s). 
 Since these photons are produced within a jet,
they come as non-isolated photons
. In this measurement, to reject the photons from mesons, the photon candidates are required to be
isolated in the calorimeter. Previous measurements of photon production at hadron colliders have
successfully used isolation-based techniques to extract the photon signal \cite{bib:prev_measurements}.


\section{CROSS SECTION MEASUREMENT}

The measurement is performed in the central detector in the kinematic range $p_T>30$~GeV and
$|\eta|<1.0$. 
The energy within a cone of radius $R=$0.4 around the photon is required to be less than 2 GeV.
The isolation
is corrected on average both for the extra energy due to transverse leakage of the photon shower and for the pile-up in the event \cite{bib:CDF_NIM}.

The trigger efficiency is measured using electrons from $Z\rightarrow ee$ data and it is 100\% in
this kinematical region
. The efficiencies of the photon selection cuts
are measured using photon Monte Carlo, and corrected by comparing data $Z\rightarrow ee$ to Monte
Carlo $Z\rightarrow ee$.




The main background contribution comes from $\pi^{0}$ and $\eta$ decays to two (or more) photons which
fake a single photon shower in the calorimeter. 
The direct photons have very limited calorimeter isolation energy, only from the underlying event,
while the fakes tend to have more isolation energy since they come inside the jet. However, occasionally these mesons
will appear isolated (when a scattered parton hadronization results in most of the energy
transferred to a single meson). These events cannot be distinguished from prompt photons on an event
by event basis. Instead, this background is removed in a statistical manner.

\begin{figure*}[t]
\centering
\includegraphics[width=90mm]{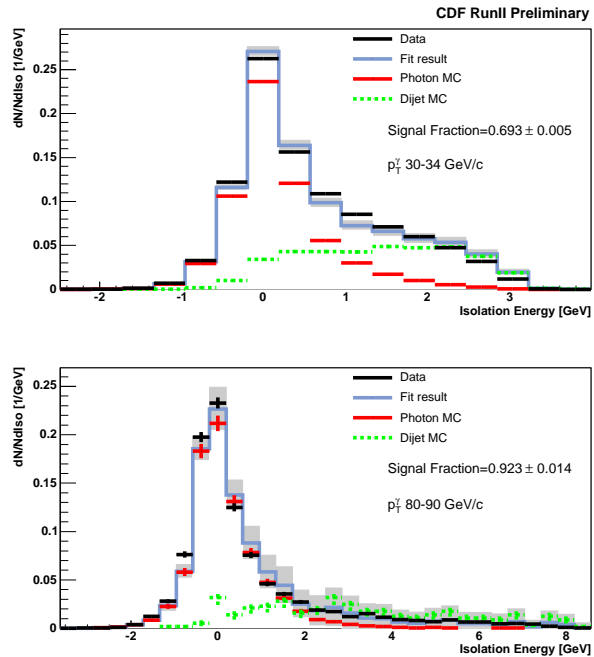}
\caption{Details of the isolation fit for two bins in $p_{T}$.\label{fig:iso_fits}}
\end{figure*}

The background subtraction method is based on the isolation energy in the calorimeter around the
photon candidate. The signal fraction is estimated by fitting the isolation distribution in the data
to signal and background Monte Carlo isolation-shape templates for every bin in photon $p_{T}$. 
The signal template is constructed with photon Monte Carlo and describes the peak of the isolation in the data. The background template
is developed from photons coming from mesons in a dijet Monte Carlo sample, and reproduces the high
isolation tail in the data (see Fig. \ref{fig:iso_fits}).

\begin{figure*}[t]
\centering
\includegraphics[width=90mm]{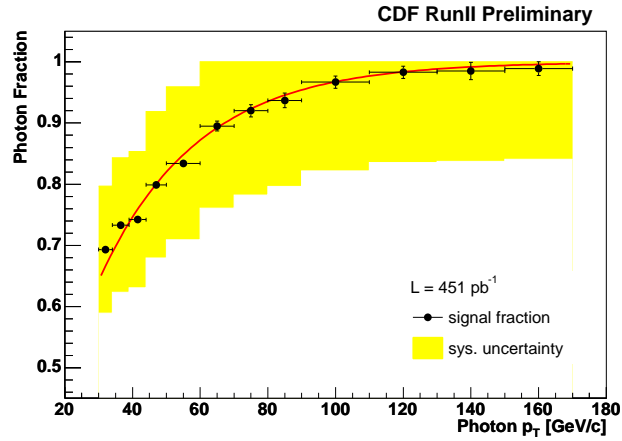}
\caption{Signal fraction as a function of the $p_{T}$ of the photon.\label{fig:signal_fraction}}
\end{figure*}

The signal fraction in our sample varies between 70\% and 100\% as the photon $p_{T}$ increases, as
displayed in Fig. \ref{fig:signal_fraction}. The
systematic uncertainty in the signal fraction (about 15\%) is determined by changing the shape of
the templates used to fit the isolation in the data.  
For the low $p_{T}$ range, instead of using photon Monte Carlo, the signal templates are constructed with electrons from $Z$
decays in data and in Monte Carlo. For the whole $p_{T}$ range, averaged-$p_{T}$ templates are used
to fit the data. This way, any $p_T$-dependence is removed from the templates. Finally, we remove
any shape consideration. Every template both in the data and in the
Monte Carlo samples is divided in two regions in isolation, above and below 2 GeV, and the signal fraction
is estimated by comparing the yields in the low isolation region to the total yield in every sample.

The measured cross section is unfolded back to the hadron level for comparing to the theory. The
unfolding factors, which take into account both 
 the efficiencies and resolution effects,  are
about 70\% and flat in $p_{T}$.

\section{RESULTS}

\begin{figure*}[t]
\centering
\includegraphics[width=90mm]{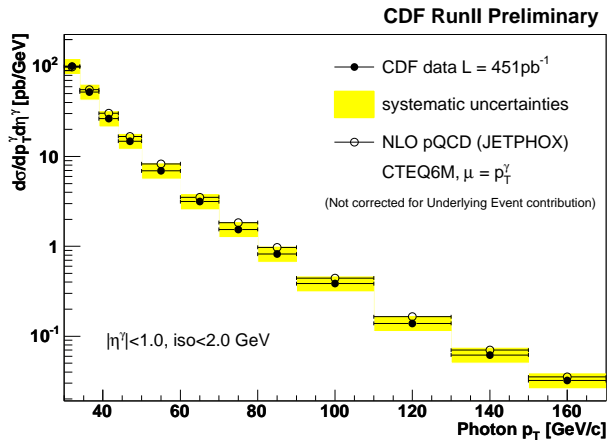}
\caption{Inclusive Isolated prompt photon cross section as a function of the $p_{T}$ of the photon.\label{fig:cross_section}}
\end{figure*}

The cross section result compared to the theory prediction is shown in
Fig. \ref{fig:cross_section}. The measurement is compared to NLO pQCD predictions from JETPHOX \cite{bib:JETPHOX}. The NLO
predictions are not corrected for non-pQCD contributions such as the underlying event. The
renormalization, factorization and fragmentation scales were chosen to be
$\mu_R=\mu_f=\mu_F=p_T^\gamma$. The CTEQ6.1M \cite{bib:CTEQ6} PDFs and the BFGII \cite{bib:BFG} parametrization are used. Data agree
with the theory within systematic uncertainties (about 20\% coming from the signal fraction, the photon
energy scale and the trigger efficiency). We expect that after the NLO pQCD
prediction is corrected for the underlying event contributions the agreement between data and theory
will improve. Future results will extend the measurement to higher transverse momenta of the photon and will
include about 3 fb$^{-1}$ of luminosity.

\end{document}